# The hidden hand of lipids in the amyloid cascade

Fabio Lolicato[1,2,♦], Carmelo Tempra[3,♦,] Martina Pannuzzo[4,♦,#] and Carmelo La Rosa[5,*]

[1]Heidelberg University Biochemistry Center, Heidelberg, Germany

[2]Department of Physics, University of Helsinki, P.O. Box 64, FI-00014 Helsinki, Finland

[3]Institute of Organic Chemistry and Biochemistry, Prague, Czech Republic

[4]Laboratory of Nanotechnology for Precision Medicine, Fondazione Istituto Italiano di Tecnologia, Via Morego 30, Genoa 16163, Italy

[5]Dipartimento di Scienze Chimiche, Università degli Studi di Catania, Viale A. Doria 6 – 95125 Catania, Italy

[♦]The authors contributed equally to this work

[#]current affiliation: Martina Pannuzzo, PhD – R&D Center, PharmCADD, Busan Republic of Korea, 48060

[*]Corresponding authors: clarosa@unict.it

## Abstract

Intrinsically disordered proteins (IDPs), such as amyloid polypeptide (IAPP), beta-amyloid (Aβ), and α-synuclein are linked to the insurgence of type 2 diabetes, Alzheimer's, and Parkinson's diseases, respectively. Common molecular mechanisms have been explored to elucidate the toxicity pathway of these proteins. For many years, the amyloid hypothesis was believed to explain the toxicity. According to this hypothesis, the misfolding of these proteins and the further aggregation into mature and insoluble fibrils rich in the β-sheet lead to cell death. However, this theory fails to explain much of the experimental evidence, which led to the hypothesis of soluble small-oligomer toxicity and pore-like activity. Recently, the lipid-chaperone model was proposed to explain the effect of lipid compositions on IDPs toxicity in vitro. In this work, we summarize the different toxicity models and discuss the possible relevance of the lipid-chaperone model in a biological context, such as protein overexpression and lipid oxidation. Furthermore, we briefly explore how the model can be incorporated into the framework that explains IDPs toxicity, such as fibril formation and secondary nucleation.

## 1. Introduction

Proteins play a crucial role in the processes that enable life. They are synthesized in the endoplasmic reticulum as a well-defined sequence of amino acids bound together by peptide bonding. After synthesis, the process of complete folding occurs in the Golgi apparatus by which proteins assume a unique three-dimensional structure. In this configuration, they can perform their biological function. However, not all proteins take on a unique three-dimensional structure; some remain random coils and are called intrinsically disordered proteins (IDPs). Among these are infamous islet amyloid polypeptide (IAPP), beta-amyloid (Aβ) fragments (39, 40 e 42 aa), and α-synuclein because they are linked to type 2 diabetes[1], Alzheimer's[2], and Parkinson's[3] diseases[4], respectively.

IAPP is a hormone-containing 37 amino acids with positions 2 and 7 linked by a disulfide bridge[5,6]. IAPP and insulin and glucagon regulate glucose concentration in the blood. Insulin and IAPP are co-secreted from islet Langharan's β cells in a ratio of 10:1. Post-mortem investigation of patients with type 2 diabetes shows that 90 % of samples have a loss of β cells and the presence of fibrils in and out of the beta cell, whereas 10 % do not show fibrils at all[1].

Aβ fragments derived from the cleavage of amyloid precursor protein (APP) are ubiquitous transmembrane proteins. Recently it was reported that amyloid oligomers of intermediate size might have a physiological relevance too: protect from infections, repair leaks in the blood-brain barrier, promote recovery from injury and regulate synaptic function[7]. Alzheimer's patients are characterized by extracellular accumulation of Aβ fragments in senile plaque and neurofibrillary tangle of hyperphosphorylated tau protein. Nevertheless, postmortem investigations have shown that, in some cases, patients with cognitive problems have no fibrillar deposits, whereas people without cognitive impairment displayed fibrillar deposits[8].

α-synuclein is a 140 amino acid protein located in the terminal of the presynaptic neurons, capable of self-assembly into toxic oligomers and fibrils. However, the biological role of α-synuclein is not clear, but it is believed to be involved in the regulation of synaptic plasticity and membrane remodeling[9]. A hallmark of Parkinson's disease is the appearance of Lewy bodies inside nerve cells consisting of α-synuclein deposits.

A common feature of these amyloidogenic diseases is that IAPP, Aβ, and α-synuclein can form aggregates toxic to pancreatic β cells and neurons. Toxicity has been associated with cell membrane disruption upon interaction with amyloidogenic proteins. Two sequential mechanisms have been discovered using model membranes interacting with amyloidogenic proteins: pore formation and detergent-like mechanism[10]. Pore formation consists of the self-assembly of proteins in the lipid phase to form isolated trans-membrane ion-channel-like pores with a diameter of approximately 1.8 nm[11–14]. Detergent-like mechanism occurs following the conversion of transmembrane assemblies into large aggregates[14], which act as detergents, tearing away lipids from the bilayer to form large breaches in the order of 500 nm in diameter[11,15]. Amyloidogenic proteins in the aqueous phase can self-assemble into unstructured aggregates called oligomers and then ordered aggregates rich in β-sheet called fibrils. The events leading to the formation of fibrils starting from the protein in its monomeric form are known as the amyloid cascade.

Some hypotheses were advanced from the above experimental fact: amyloid, toxic oligomers, and lipid-chaperone.

## 2. The Amyloid Cascade Hypothesis

Originally the term amyloid cascade and amyloid hypothesis were synonymous. However, after 25 years of studies on amyloidogenic proteins and an impressive amount of data, the term amyloid cascade has been enriched with new concepts such as primary and secondary nucleation of fibrils, formation of oligomers and other pre-fibrillar species, pore formation, detergent-like and carpeting, and membrane damage. In the following paragraphs, we discuss these aspects divided into the amyloid hypothesis, the toxic oligomer hypothesis, and a more general hypothesis that combines the amyloid and toxic oligomer hypothesis into a single framework, the lipid-chaperone hypothesis.

## 2.1 The Amyloid Hypothesis

The idea of the amyloid hypothesis was born in 1984 and is due to George Glenner, who, together with Caine Wong, proposed that Alzheimer's disease is due to the accumulation of Aβ in the brain[16]. This accumulation results in the loss of cognitive abilities of people with Alzheimer's disease. This idea has attracted much criticism and skepticism, although much clinical evidence supports the amyloid hypothesis[17,18]. On the other hand, the amyloid hypothesis fails to explain why some patients who do not show fibrillar deposits have cognitive problems, while some patients who do not have fibrillar deposits have evident cognitive impairments[8]. Moreover, these kinds of aggregates have been observed in other devastating diseases, including Parkinson's disease, type 2 diabetes mellitus, amyotrophic lateral sclerosis, prion disease, hereditary lysozyme amyloidosis, and many others[19].

Biophysical investigations have shown that the formation of fibrils is a very complex phenomenon involving several steps starting from proteins in monomeric form until they reach the stable form called fibrils: pseudo ordered proteic aggregate rich in β-sheet. Fibrils are formed through a two nucleation step self-assembling phenomenon, which leads to structural variants as shown by SS-NMR, X-ray, and cryo-microscopy investigations[20,21]. Indeed, X-ray investigations revealed a diffractogram characterized by an intense and sharp ring at 4.75 Å, an overlapped and diffuse one at 4.3 Å, and a broad and less intense ring centered at 9.8 Å[22,23]. Fibrils derived from amyloidogenic proteins are resistant to digestion with proteinase K[24]. In addition, it was observed that fibrils obtained in vitro and ex-vivo are structurally different[25].

Although the amyloid hypothesis was developed for Alzheimer's disease, this concept has been extended to other diseases such as Parkinson's disease, tauopathies, and type 2 diabetes. These diseases share many common features in *in vitro* systems, from a molecular point of view[11,19,26–30]. They form fibrils, ion-channel-like pores, and damage model membranes by detergent-like or carpeting mechanisms.

Developing drugs to target soluble oligomers and amyloid plaques is a good test of whether the amyloid hypothesis interprets the above evidence well.

A reasonable strategy in drug development is to consider the disease as a sequence of chemical reactions starting from reactants (the initiation) that lead to products (the end). Then, having determined all the steps in this reaction sequence, one can choose a particular step and design and synthesize an inhibitor for that step. Finally, if the hypothesis is correct (and the drug works), the reaction is stopped, and the products are not reached.

However, all drugs in phase III clinical trial tests produced so far lack efficacy. They are all based on the amyloid hypothesis, i.e., the target of these drugs have been soluble oligomers, amyloid plaques, and soluble Aβ[31]. From the results of clinical trials, it can be concluded that the amyloid hypothesis might not reflect the pathology of the disease, so it is reasonable to look for other routes.

**2.2 The Toxic Oligomers Hypothesis**

The amyloid hypothesis predicts that the cause of IDP-induced diseases is due to the accumulation of fibrils in the various organs involved. More recently, Goldberg and Lansbury proposed the toxic oligomer hypothesis according to which, rather than the accumulation of fibrils, the cause of the disease is due to the accumulation of oligomers[32].

The self-assembling pathway of amyloidogenic proteins (**Figure 1, points 1-5**) starts from their respective monomers and leads to the formation of unstructured and transient oligomers, followed by the formation of pre-fibrillar species. Then, the primary nucleation of non-branched fibrils occurs, succeeded by the final elongation and secondary nucleation, which leads to the formation of branched fibrils. Oligomers are transient soluble species that are noncompact and poorly structured. They are formed at the beginning of the amyloid cascade and by the dissociation of fibrils formed in the secondary nucleation[33]. According to recent reports, the latter are more toxic to cells than those formed at the beginning of the amyloid cascade[28,34,35]. It has been observed that oligomers formed by IAPP, α-synuclein, and Aβ share many chemical-physical and biological properties.

Moreover, IAPP, Aβ, and α-synuclein can form mixed oligomers between them[1,36,36,37]. Since toxicity is associated with the membrane damage done by amyloidogenic proteins, it is more interesting to discuss oligomers' effect in interacting with the membrane. Since the cell is a very complex system, model systems have been widely used following the bottom-up investigation in studying the interaction of amyloidogenic proteins and cell membranes. In particular, since their chemical and chemical-physical characteristics have been well characterized, phospholipid vesicles (e.g., LUV and GUV) are among the most used to mimic biological membranes[38–43]. The oligomers formed by α-synuclein, Aβ, and IAPP have many behaviors in common in interaction with model membranes. IAPP, α-synuclein and Aβ oligomers form, in a first step, small pores permeable to ions; then, in a second step, the membrane is damaged by a detergent-like process[11,26,30,44–47].

## 2.3 The Lipid Chaperone Hypothesis

In the previous chapter, we have pointed out the importance of membrane-mimicking organelles such as GUVs or LUVs for studying the toxicity of IDPs. The lipid chaperone model[44,48–51] was initially developed to explain the effect, in vitro, of different lipids on IDPs toxicity. The model is based on the thermodynamic equilibrium of lipids between the water phase and lipid-organelles such as GUVs, LUVs and any kind of membrane. This equilibrium is defined as Critical Micellar Concentration (CMC) and the lipids in the water phase are named free lipids from now on. The free lipids can interact with the IDPs to form a protein-lipid complex. To briefly explain the model, these reactions should be taken into account:

$$M_n \rightleftarrows nL(aq) \quad (a)$$

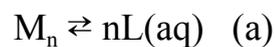

$$L + P \rightleftarrows LP \quad (b)$$

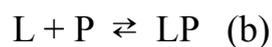

$$P + P_{n-1} \rightleftarrows P_n \quad (c)$$

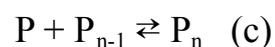

Where $M_n$ is a generic membrane organelle (micelle, LUV, etc.), $L$ is a single lipid in the aqueous phase. $P$ is a protein monomer in a random coil conformation, $P_n$ and $P_{n-1}$ are fibrillar aggregates and $LP$ is the protein-lipid complex. The equilibrium (a) is defined as CMC. The equilibrium (c) is the classical aggregation pathway to form fibrillar aggregates rich in β-sheet.
The equilibrium constant of the specie LP will be:

$$K_{LP} = \frac{[LP]}{[L][P]} \quad (eq\ 1)$$

and Gibbis free energy of lipid-protein complex can be defined by the following:

$$\Delta G = -RT\ln K_{LP} \quad (eq\ 2)$$

Where **R** is the gas constant and **T** the temperature. **LP** species are formed in excess compared to **L** or **P** only when Gibbs free energy is <0, so **K**>1. Furthermore, the equilibrium (b) can be shifted to the right by increasing the reactant concentration, i.e., lipid (e.g., lipid oxidation) or protein concentration (protein overexpression).

According to the lipid-chaperone model, the species **LP** is more prone to insert and damage the membrane through pore formation. For this to happen, the protein-lipid complex has to be more hydrophobic than the bare protein.

When the concentration of free lipids (**L**) is high enough, equilibrium (b) is shifted to the right, while equilibrium (c) is shifted to the left. The result is that the aggregation into fibrillar species is inhibited, while the channel-like toxicity is enhanced both from a kinetic and thermodynamic point of view. The concentration of **L** is dictated by the CMC, meaning that a change in the CMC will change the equilibrium and toxicity effects discussed above. Furthermore, by combining spectroscopic experiments and molecular dynamics simulations, we revealed a conformational transition when the free lipids interact with the protein. In fact, without free lipids, the protein's conformation changes from a random coil to a β-sheet-rich structure when forming fibrils; with free lipids, the conformation of the protein changes from a random coil to alpha-helix while increasing the channel-like activity. The α-helix region seems to be responsible for the binding with amphiphilic molecules fostering the proposed mechanism[52].

Well documented is, for example, the propensity of a α-synuclein to assume an α-helical conformation upon binding to docosahexaenoic acid (DHA), one of the fatty acids found in high levels in the brain of PD patients[53].

From the chemical equilibrium (b) described above, it can be deduced that the formation and the nature of the amyloidogenic lipid-protein complex depend on the concentration of the free lipid and protein in the aqueous phase. Therefore, one can shift the equilibrium to the right (complex formation) as the concentration of the phospholipid or protein increases to reach the minimum free energy (eq 2) compatible with the considered system. The lipid-protein complex's formation triggers the amyloidogenic protein's insertion into the lipid bilayer and, subsequently, the formation of ion-channel-like pores. This model suggests that amyloidogenic protein toxicity does not only depend on protein overexpression but also on the presence of free lipids in the aqueous phase.

From the above, it is clear that these processes are subject to chemical equilibrium; besides, life itself is based on chemical equilibrium.

## 3.0 Perspectives

The Lipid-chaperone model was initially developed to explain the effect of different lipids (with different CMC) on the competition between fibril growth and pore formation. However, cell membranes are complex systems with properties such as membrane fluidity, bending rigidity, and lipid diffusion, which differ from simple one-component model membranes. Furthermore, lipophilic molecules other than phospholipids are present in cell membranes, such as fatty acids and detergents, and they could serve as chaperones for the IDPs. Moreover, free lipids in equilibrium between the water phase and lipid-organelles might not be the unique source of available lipophilic molecules, e.g., some amyloid proteins are known to be transported on lipoproteins[54]. With this chapter, we want to go beyond our model, suggesting future directions to fill the gaps between simple in vitro biophysical models and more complex cell membranes.

## 3.1 Membrane Composition

The current view of cell membrane structure hypothesizes that the cell membrane is a lipid bilayer, in which proteins are embedded and carry out most of the membrane functions[55]. Lipids and proteins are responsible for the membrane's fluidity and undergo lateral and rotational diffusion. The central components of cell membranes are phospholipids. These lipid molecules comprise a (hydrophilic) phosphate-based head group and two (hydrophobic) fatty acid chains. Another essential lipid is cholesterol, which regulates the cell membrane's fluidity and permeability[56–58]. Nobody knows how many different lipid types there are in cell membranes since this number likely depends on cell type, diet, and many other factors, but there are many, of the order of a hundred. Due to the cell membrane complexity, it is clear that simple (one component) model membranes are too simplistic to mimic the properties of the cell environment. Therefore, we believe the lipid-chaperone hypothesis should be tested in vitro with a more realistic lipid composition and cholesterol content. Indeed, lipid asymmetry, cholesterol content, and a crowded protein environment are known to affect membrane properties such as membrane fluidity, lipid diffusion, and membrane rigidity, as well as amyloid protein aggregation[56,59,60]. Could they also affect the CMC? Could it mean that the free lipid concentration in an aqueous solution can be influenced by these parameters and regulate the switch between ion-channel-like formation, detergent-like mechanism, and fibril formation?
Furthermore, using NMR and MD simulations, it has been shown that the lipid-protein complex exists[44], confirming the lipid-chaperone hypothesis in vitro. However, it is still unclear which of these lipids is responsible in a cellular context. In addition to phospholipids and cholesterol, cell membranes contain various other lipids such as sterols, sphingolipids, fatty acids, and lipid derivatives, as well as carbohydrates anchored to proteins and lipids on the extracellular leaflet. Any of those could potentially interact and form a complex with amyloidogenic proteins[61]. The next step will be to systematically estimate the binding affinity between IDPs and each lipid class to see if we can understand protein's preferentiality and if it changes with different amyloid proteins, lipid environments, or cholesterol concentrations. If this is the case, the protein-membrane interaction with a specific organelle can be driven by the presence of a particular type of lipid or cholesterol content.

We believe that future efforts to widen the lipid-chaperon hypothesis should be soon made in this direction by the scientific community.

## 3.2 The role of protein concentrations

The production of amyloidogenic proteins is physiological, and their local concentration can vary depending on the needs. Under physiological conditions, Aβ concentrations vary according to neuronal activity. For example, at picomolar concentrations, Aβ induces long-term hippocampal potentiation (LTP), a form of synaptic plasticity probably implicated in learning and memory. In contrast, nanomolar Aβ concentrations inhibit it[62].

Upon neuronal insults, Aβ$_{42}$ oligomers induce an increase of Ca$^{2+}$ in presynaptic terminals responsible for suppressing synaptic vesicle transport via activation of the CaMKK-to-CaMKIV pathway[63]. This way, Aβ$_{42}$ oligomers prevent the formation of new synapses and may promote neuronal survival under challenging conditions. The Aβ peptide may thus protect neurons from damage by serving as a negative feedback mechanism to prevent excessive synaptic activity and excitotoxicity[7].

Amylin concentrations, instead, oscillate in healthy humans in response to glucose and free fatty acids[64]. Plasma amylin levels can exceed that of insulin during a persistent hyperglycemic state. Overexpression of IAPP inhibits insulin secretion and could protect islets under metabolic stress[65].

IAPP affects bone metabolism like the homologous calcitonin gene-related peptide (CGRP)[66]. Amylin crosses the blood-brain barrier (BBB)[67] and, similarly to A$\beta_{42}$, at a low dose, can induce depression of LTP[68] and mediate an increase of cytosolic cAMP and $Ca^{2+}$ [69].

By contrast, α-syn is localized in the presynaptic terminal of neurons and seems to regulate vesicle dynamics. The release of endogenous α-syn depends on neural activity[70]. Upon neuronal stimulation and calcium influx, α-syn promotes vesicular trafficking and neurotransmitter release[71]. It has been shown to physiologically control mitochondrial $Ca^{2+}$ homeostasis, mediating, in turn, the autophagic process[72].

The transiently increased levels of amyloid peptides associated with the cytosolic increase of $Ca^{2+}$ ions can thus be functional in healthy conditions[72,73]. Less clear is the molecular mechanism by which amyloid peptides induce $Ca^{2+}$ currents. As evidenced elsewhere, amyloid peptides can directly interact and activate $Ca^{2+}$ channels or change the membrane's mechanical properties, stimulating the activity of transmembrane $Ca^{2+}$ channels[74–77]. It has also been proposed that amyloid peptides may form functional channels to mediate the calcium transport across membranes[7]. The hypothesis comes from the experimental observation that amyloid oligomers share the common ability to form $Ca^{2+}$ permeable pores in the membrane spontaneously[45].

Might, then, exist a protein concentration threshold above which the formation of transmembrane pores becomes toxic, whereas below which ensure the physiological influx of $Ca^{2+}$ ions inside the cell?

**3.3 The role of lipid concentrations**

The lipid-chaperone hypothesis is based on free lipids that amyloidogenic proteins can sequester from the chemical equilibrium with the vesicles. Stable protein-lipid complexes are transported into the bilayer due to their higher hydrophobicity. This idea is supported by membrane leakage experiments on three different amyloidogenic proteins as a function of lipids having different CMC. Furthermore, the formation of the lipid–IDP complexes was confirmed by numerous experimental techniques such as 2D NMR, circular dichroism (CD) spectroscopy, MD simulations, and isothermal titration calorimetry (ITC) measurements[44]. The data demonstrate the existence of a stable complex with increased hydrophobicity, essential prerequisites to lipid-assisted protein transport.

A question always arises from biochemistry and cell biology scientists if there could be additional sources of lipids rather than the free lipids coming from the chemical equilibrium with the vesicles and how it is related to pathological conditions.

In physiological conditions, free phospholipids are usually removed from the aqueous phase by the phospholipase A1 enzyme activity that captures and transforms free lipids into lysophospholipids and fatty acids[78]. This process is in a dynamical equilibrium with phospholipid synthesis, mainly occurring in the endoplasmic reticulum. Perhaps in pathological conditions, e.g., due to a miss function of phospholipase enzymes, this equilibrium might be more shifted towards forming free lipids. If this occurs, amyloidogenic IDPs could sequester free lipids more easily, forming stable complexes more frequently and initiating the amyloid cascade. Interestingly, a higher concentration of short-chain phospholipid (high CMC) was found in neurons of Alzheimer's patients than in healthy people[79].

Oxidized phospholipids could serve as an alternative source of free lipids. Interestingly, oxidized phospholipids have already been shown to play a crucial role in type II diabetes, Alzheimer's, and Parkinson's deseases[80–89]. Furthermore, they also have a two order of magnitude higher CMC[90,91] compared to the non-oxidized phospholipid, meaning they can foster pore formation based on the lipid-chaperone hypothesis. Like the phospholipase A1 enzyme, in physiological conditions, oxidant machinery, e.g., superoxide dismutase-catalase enzyme, continually destroys reactive oxygen species (ROSs) formed during molecular oxygen exchange. However, under pathological conditions, redox homeostasis is unbalanced, producing an excess of lipid peroxides[89], which could serve as a source of free lipid for IDPs.

Another largely unclear aspect is how amyloidogenic proteins are transported and metabolized once released from the cell and if they are associated with other proteins in the bloodstream or exist as free proteins.

Amyloid peptides are highly fibrillogenic and strongly interact with lipids and membranes physiologically and pathologically[92]. As these peptides are known to travel between different organelles, cells, and tissues, in physiological conditions are probably protected from spontaneous aggregation or unspecific interactions by forming a complex with other components. For example, several groups reported Aβ binding to Apolipoprotein E (ApoE) and High-density lipoprotein (HDL) particles under in vitro conditions[54,93–97], suggesting a role in the trafficking and clearance of Aβ[98].

Accumulating amyloid peptides might result from abnormalities in this protective mechanism due to genetic mutations or environmental stress. The isoform ApoE4, associated with the onset of Alzheimer's disease, has a lower lipid-loading capacity and reduced Aβ-binding affinity compared with the other two isoforms, ApoE2 and ApoE3. A reduced affinity between ApoE4 and amylin correlates with increased amyloid nucleation and aggregation propensity[99]. A recent study has evidenced that ApoE4 disrupts microglia's lipids catabolism, with an evident reduction of lipid export and re-uptake, resulting in lipid intracellular accumulation. Reduced binding of ApoE4 to amyloidogenic peptides does not prevent amyloid peptides from binding to lipids, but their trafficking will likely be affected. It is worth noting that the distribution of polar and apolar residues in α-syn reminds the amphipathic α-helices motifs common to apolipoproteins and other lipid-binding proteins[100]. In turn, the lipid-binding motif in apolipoproteins has shown to have a high amyloidogenic propensity[101].

It follows that soluble amyloid peptides are likely associated with lipid and/or transport-binding counterparts that can become partners in crime in challenging circumstances. Notably, fatty acid (FA) abundance and lipid metabolism are also altered in AD pathology[102,103]. Elevated levels of circulating free fatty acids, as evidenced in pre-diabetic or T2DM patients, are probably involved in amyloid formation[104].

α-syn toxicity and PD progression are associated with dysregulated lipid metabolism[105].

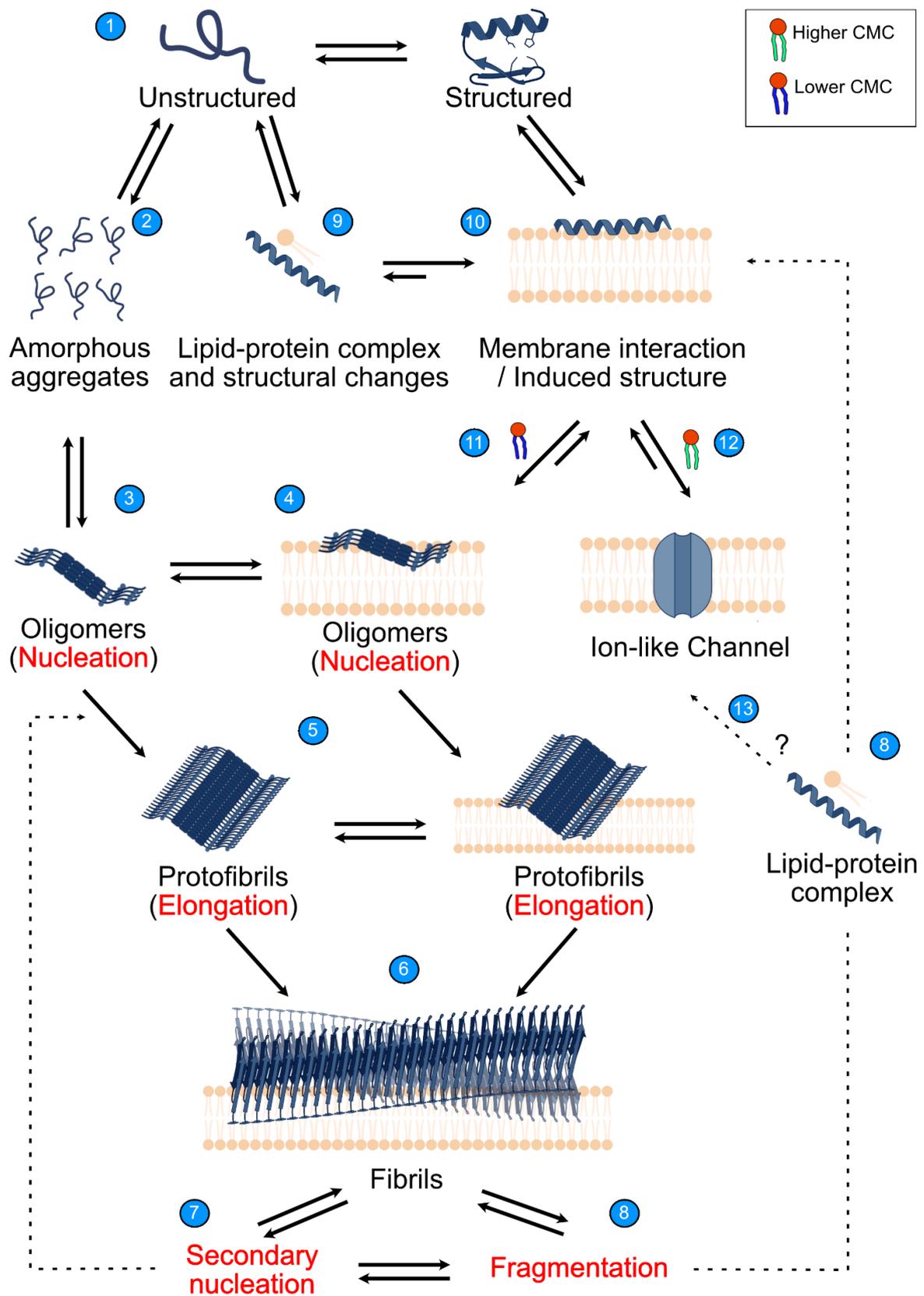

**Figure 1**. **A unifying framework for the self-assembling pathway of amyloidogenic proteins.** Created with BioRender.com

## 4. Discussion

IAPP, Aβ, and α-synuclein form aggregate toxic to pancreatic β cells and neurons. The most accredited hypothesis to date foresees toxicity associated with cell membrane disruption upon interaction with amyloidogenic proteins. Nevertheless, the molecular mechanism is still unknown, and there is still no unique theory that can explain some facts, for example, at the same time why fibrils are found in healthy patients without any cognitive issues, whereas others develop Diabetes type 2, Alzheimer and Parkinson's without fibrils.

We are now proposing a unifying framework for amyloid-mediated membrane damage to explain these divergences (**Figure 1**). If our model is correct, we should observe only plaque formation in dementia-suffering patients if lipid grafting occurs after plaque accumulation. However, If lipid grafting occurs earlier, i.e., without protein overexpression, it could explain the onset of dementia without plaques. Plaques should not be found if there is no lipid grafting, and no dementia should occur.

As shown in **Figure 1, point 1**, amyloidogenic intrinsically disordered proteins are a particular class of proteins. Their secondary structure is dynamic, in equilibrium between unfolded, partially folded, and fully folded states. This equilibrium can be influenced by numerous variables such as pH, ionic strength, salt type, or the binding with other biomolecules.

The classical pathways leading to fibril formation provide unfolded or partially folded monomers able to form amorphous aggregates (**point 2**). These aggregates can evolve into structured oligomers able to bind to cellular membranes (**points 3-4**). Aggregates act as nucleation seeds for the formation of protofibrils first (**point 5**) and fibrils later (**point 6**).

Fibrils, at this stage, can undergo secondary nucleation (**point 7**), which increases the rate of formation of low–molecular weight oligomers (**point 8**) that accompany the formation of mature fibrils (**point 7**)[33,106].

The presence of a lipid-protein complex, resulting from the interaction of an unstructured oligomer with free lipids, could lead to an alternative path for fibril growth and membrane damage. Indeed, as proposed by the lipid-chaperone hypothesis, the formation of lipid-protein complex induces a secondary structure transition from unstructured to alpha-helix (**point 9**).

The protein-lipid complex has a hydrophobicity higher than the bar protein and will be more prompt to interact with the membrane[48] (**point 10**).

Now, depending on the lipid nature (CMC), the protein-lipid complex can begin the fibril growth (**point 11**) or the formation of an ion-like channel (**point 12**).

An unknown aspect is whether protein fragments derived by fragmentation might be already structured oligomers with a predominance of β-sheets (since fibrils are mostly β-sheets). We think that these second-generation low–molecular weight structured oligomers derived from fragmentation could also be associated with lipids (**point 8**). For example, membrane adsorbed fibrils might damage the membrane by scratching lipids from its surface, fostering the interaction with fibril fragments. It will be interesting to study the nature of their structure and if they can make a complex with lipids that can promote their insertion into the membrane, leading to toxic *β*-barrel ion-like channels, as suggested in **point 13**.

## 5. Conclusions

Type 2 diabetes, Parkinson and Alzheimer's are multifactorial illnesses, so many factors can trigger pathologies. The above discussion emerges the concept that the binding of amyloid peptides to spare lipids triggers toxicity; the possible source of free lipids should be oxidized phospholipids (dyshomeostasis of anti-oxidant machinery) or phospholipase enzymes capable of converting di-acyl-phospholipid into lysophospholipids or ApoE chaperone dyshomeostasis.

IAPP, A$\beta$ and $\alpha$-synuclein can all form ion-channel-like pores spontaneously in model membranes. We can postulate the existence of functional and toxic ion-channel-like pores formed by amyloidogenic proteins. As with everything in life, all those states are equilibrium processes where the concentration of lipid or protein as well as external biomolecules could foster one or the other direction. For example, our data showed that in the presence of short-chain lipids, r-IAPP could form membrane pores[44], suggesting that this class of proteins might also create functional pores in physiological conditions, e.g., for $Ca^{2++}$ transport in/out cell compartments.

However, three questions arise: i) can the discriminant between functional (coming from the expression of protein) and toxic pores (coming from fragmentation) be the protein secondary structure, i.e., $\alpha$-helix functional and $\beta$-sheet toxic?

ii) Can functional pores become toxic if their concentration increases, unbalancing the physiological ions influx between cell compartments, e.g., due to a miss function of phospholipase enzymes or dyshomeostasis of anti-oxidant machinery?

iii) Can specific lipid types and their concentrations tune the functional-toxic behavior of amyloidogenic IDPs by inducing secondary structural changes or disturbing the physiological $Ca^{2+}$ signaling?

In conclusion, we believe that the determining steps leading to toxicity are lipid-protein complex formation, fragmentation, and secondary nucleation (*β*-sheet structure oligomers reserve), with the cellular lipid environment playing a crucial role. Therefore, drug development should be aimed at blocking these reactions, and future studies should systematically explore lipidomics changes with pathology, aging, and at different disease stages.

## Acknowledgment

We thank every one of our collaborators and colleagues with whom we have discussed this work during the many years. All the authors are grateful to Professor Antonio Raudino, who passed away in 2021, for his contribution to developing the Lipid-chaperone hypothesis. C.T. would like to acknowledge the International Max Planck Research School for Many-Particle Systems in Structured Environments hosted by the Max Planck Institute for the Physics of Complex Systems, Dresden, Germany. F.L. gratefully acknowledges the data storage service SDS@hd supported by the Ministry of Science, Research and the Arts Baden-Württemberg (MWK) and the German Research Foundation (DFG) through grant INST 35/1314-1 FUGG and INST 35/1503-1 FUGG. This article is partially supported by the University of Catania grant "*PiaCeri linea di intervento 2*". CSC–IT Center for Science (Espoo, Finland) is thanked for the ample computing resources that have rendered our research possible.